# Characterizing the zenithal night sky brightness in large territories: How many samples per square kilometer are needed?

**Salvador Bará** [1,*]

[1]*Departamento de Física Aplicada, Universidade de Santiago de Compostela, 15782 Santiago de Compostela, Galicia, Spain.*

[*] *salva.bara@usc.es*

## Abstract

A recurring question arises when trying to characterize, by means of measurements or theoretical calculations, the zenithal night sky brightness throughout a large territory: how many samples per square kilometer are needed? The optimum sampling distance should allow reconstructing, with sufficient accuracy, the continuous zenithal brightness map across the whole region, whilst at the same time avoiding unnecessary and redundant oversampling. This paper attempts to provide some tentative answers to this issue, using two complementary tools: the luminance structure function and the Nyquist-Shannon spatial sampling theorem. The anaysis of several regions of the world, based on the data from the *New world atlas of artificial night sky brightness* (Falchi et al 2016, Sci. Adv. 2, e1600377) suggests that, as a rule of thumb, about one measurement per square kilometer could be sufficient for determining the zenithal night sky brightness of artificial origin at any point in a region to within ±0.1 mag$_{\rm v}$/arcsec$^2$ (in the root-mean-square sense) of its true value in the Johnson-Cousins V band. The exact reconstruction of the zenithal night sky brightness maps from samples taken at the Nyquist rate seems to be considerably more demanding.

## 1. Introduction

The night sky brightness in vast regions of the world is significantly higher than its expected natural value (Cinzano et al. 2005; Kyba 2015; Bará 2016; Falchi et al. 2016, 2016b), due to atmospheric scattering of visible photons emitted by artificial light sources. This is one of the most conspicuous manifestations of light pollution, a generic term that encompasses the unwanted consequences of the present way of using artificial light. Significant efforts have been devoted in recent years to model and quantify this phenomenon (Garstang 1986; Cinzano et al. 2001; Cinzano & Elvidge 2004; Kocifaj 2007; Cinzano & Falchi 2012; Kyba et al. 2012; Aubé 2015; Aubé et al. 2016; Kocifaj 2016; Ribas et al. 2016; Solano-Lamphar & Kocifaj 2016), as well as to evaluate the effects of artificial light at night on the environment (Longcore & Rich 2004; Rich & Longcore 2006; Navara & Nelson 2007; Hölker et al. 2010, 2010b; Gaston et al. 2013, 2014).

Several radiometric magnitudes can be used to characterize the night sky brightness (NSB), and to evaluate its departure from the expected natural conditions at any given site. The most comprehensive description is provided by the spectral radiance (in $Wm^{-2}sr^{-1}nm^{-1}$), specified for all directions of the sky hemisphere above the observer and for all wavelengths of the optical spectrum. The measurement of the spectral radiance, however, requires using relatively complex and expensive equipment (e.g. sequential spectrometers or all-sky hyperspectral cameras), and this has precluded its generalized use in field conditions. A techically simpler but still highly useful approach is provided by all-sky imaging in the Johnson-Cousins B, V and R bands (Duriscoe et al. 2007; Rabaza et al. 2010; Aceituno et al. 2011; Falchi 2011), or by conventional DSLR RGB imagery (Kolláth 2010, Jechow et al. 2017). The all-sky photopic brightness evaluated in the $V(\lambda)$ band or, as an approximation, in the Johnson-Cousins V, allows one to calculate a relevant set of visual parameters at the observer's site (Duriscoe 2016).

The zenithal NSB, despite having a considerably smaller information content than the all-sky distributions described above, is widely used nowadays to characterize the anthropogenic disruption of the natural night. This is partly due to the widespread availability of low-cost detectors that enable its straightforward measurement, by professional and citizen scientists alike, in many countries of the world (Pun & So 2012; Espey & McCauley 2014; Puschnig et al. 2014, 2014b; den Outer at al. 2015; Kyba 2015; Bará 2016; Zamorano et al. 2016; Ges et al., 2017). Mapping the zenithal NSB in large territories is interesting for a wide range of studies, including the validation of numerical models of light pollution propagation through the atmosphere. Several research groups have measured the zenithal night sky brightness in dense grids across extended regions, in order to elaborate continuous light pollution maps (Spoelstra & Schmidt, 2010; Fersch & Walker, 2012; Ribas et al., 2015; Sánchez de Miguel, 2016; Ribas, 2017; Zamorano et al., 2016).

The zenithal NSB at any observing site is not a fixed number, but a highly variable physical magnitude that depends on the changing state of the atmosphere (most notably on the aerosol concentration profile and on the presence and type of clouds), on the deterministic evolution of the emissions from artificial light sources located in a region that may be hundreds of km wide, and on the natural sources located above the observer (celestial objects in the zenithal region of the sky, and natural airglow). Under a layered atmosphere with constant conditions the zenithal NSB varies from site to site due to the change in the relative position of the observer with respect to the artificial sources, as well as to the particular distribution of obstacles that may block the atmospheric light propagation along certain paths. The zenithal NSB at neighbouring locations, however, tends to be partially correlated, since the scattering at any elementary volume of the air column above each observer involves multiple superposition integrals of the radiances emitted by a shared set of artificial sources. The question arises, then, of what is the maximum permissible distance between adjacent measurements in a given territory, in order to ensure that after a suitable mathematical processing a reasonably accurate reconstruction of the true zenithal night sky brightness can be achieved for every location.

Some tentative answers to this issue are explored in this paper, based on the analysis of the luminance structure function and on the spatial version of the Nyquist-Shannon sampling theorem, whose definition and properties are briefly described in Section 2. To get some insight about the expected order of magnitude of the optimal sampling distance, both methods are applied in Section 3 to the zenithal NSB distributions in several regions of the world, whose artificial component has been estimated by Falchi et al. in *The new world atlas of artificial night sky brightness* (Falchi et al. 2016, 2016b), henceforth referred to as the NWA. The analysis performed for the artificial component of the zenithal NSB in Section 3.1. is extended to the total one in Section 3.2., and the differences and similarities between the two cases are pointed out. As a result, some practical consequences can be extracted regarding the optimum sampling distance under a variety of situations. The significance and limitations of this study are addressed in Section 4, and Conclusions are drawn in Section 5.

## 2. Methods

The methods described in this section can be equally applied to the artificial component of the night sky brightness or to the total, including the contribution of natural sources such as celestial bodies and atmospheric airglow. Quantitative results for each case will be presented in Section 3.

### *2.1. The zenithal sky brightness spatial structure function*

Let $L(\mathbf{r})$ be the zenithal night sky brightness recorded by an observer located at the point $\mathbf{r}$ in a given geographical region. "Sky brightness" is a short-hand term for the integral over wavelengths of the spectral radiance at the entrance of the detector, weighted by the spectral filter function of the photometric band in which the observations are carried out, e.g. the Johnson-Cousins V (Bessell 1979), or the CIE scotopic V'($\lambda$) or photopic V($\lambda$) bands (CIE 1990). When the weighting function is the spectral efficacy of the human visual system for the appropriate level of luminance adaptation (photopic or scotopic), the resulting sky brightness can be expressed in SI luminance units of cd/m$^2$, equivalent to lx/sr. Otherwise, and according to the SI recommended practice, its value is given in weighted Wm$^{-2}$sr$^{-1}$, specifying the measurement band.

The NSB has a wide dynamic range, and it is often conveniently expressed in the negative logarithmic scale of magnitudes per square arcsecond (mag$_V$/arcsec$^2$). The weighted radiance $L$ in the Johnson-Cousins V band and its value $m$ in mag$_V$/arcsec$^2$ can be approximately related by (Bará 2016):

$$L[\mathrm{Wm}^{-2}\mathrm{sr}^{-1}] = 158.1 \times 10^{(-0.4m)},$$ (1)

as can be deduced from the associated SI luminance, in cd/m$^2$, which is conventionally estimated as (Garstang 1986; Kyba 2015; Bará 2016; Sánchez de Miguel et al. 2017)

$$L[\mathrm{cd}\cdot\mathrm{m}^{-2}] = 10.8 \times 10^4 \times 10^{(-0.4m)}.$$ (2)

Eq. (1) results from dividing Eq. (2) by the 683 lm/W scale factor that accounts for the maximum luminous efficacy of the optical radiation for photopically adapted eyes. Note, however, that Eqs. (1) and (2) are only approximate, because the Johnson-Cousins V and the CIE photopic V($\lambda$) bands are not strictly equivalent (Sánchez de Miguel et al. 2017).

When moving from $\mathbf{r}$ to a neighbouring place $\mathbf{r}' = \mathbf{r} + \mathbf{d}$, the zenithal NSB $L(\mathbf{r})$ changes to $L(\mathbf{r}')$. The difference $L(\mathbf{r}+\mathbf{d}) - L(\mathbf{r})$ generally depends on $\mathbf{r}$ and $\mathbf{d}$, and shall be evaluated on a case by case basis. However, some insights about its expected behaviour can be obtained by computing the spatial average of its squared value over an extended area, $S$. The result is the *zenithal sky brightness spatial structure function*, defined in energy or luminance units as

$$D_L(\mathbf{d}) = \left\langle [L(\mathbf{r}+\mathbf{d}) - L(\mathbf{r})]^2 \right\rangle = \frac{1}{S} \iint_S [L(\mathbf{r}+\mathbf{d}) - L(\mathbf{r})]^2 \, \mathrm{d}^2\mathbf{r},$$ (3)

where $\mathrm{d}^2\mathbf{r} = \mathrm{d}x\mathrm{d}y$ is the surface element in $S$, and the brackets denote spatial averaging. If the zenithal brightness data are available as an array of finite-sized spatial pixels, the integral in Eq. (3) becomes a finite sum. Note that Eq. (3) can equivalently be written as:

$$D_L(\mathbf{d}) = \left\langle [L(\mathbf{r} + \mathbf{d})]^2 \right\rangle + \left\langle [L(\mathbf{r})]^2 \right\rangle - 2\langle L(\mathbf{r} + \mathbf{d})L(\mathbf{r}) \rangle, \qquad (4)$$

whose last term is the *zenithal sky brightness spatial correlation function* $B_L(\mathbf{d}) = \langle L(\mathbf{r} + \mathbf{d})L(\mathbf{r}) \rangle$. In the limit of very large averaging areas (i.e., those whose linear dimensions are very large compared with the modulus of $\mathbf{d}$), we have $\left\langle [L(\mathbf{r} + \mathbf{d})]^2 \right\rangle = \left\langle [L(\mathbf{r})]^2 \right\rangle = \sigma_L^2$, where $\sigma_L^2$ is the mean squared value of $L(\mathbf{r})$ in $S$. Consequently, for large $S$, the spatial structure and correlation functions are related by

$$D_L(\mathbf{d}) = 2\sigma_L^2 - 2B_L(\mathbf{d}). \qquad (5)$$

The spatial structure function can also be defined for the zenithal night sky brightness expressed in mag$_V$/arcsec$^2$ units, $m(\mathbf{r})$, as

$$D_m(\mathbf{d}) = \left\langle [m(\mathbf{r} + \mathbf{d}) - m(\mathbf{r})]^2 \right\rangle. \qquad (6)$$

Equations similar to Eqs. (4)-(5) immediately follow, after substituting $m$ for $L$.

The expected change of the zenithal sky brightness as the observers move from their initial observing place $\mathbf{r}$ to a new position $\mathbf{r}' = \mathbf{r} + \mathbf{d}$ can be estimated (in the rms sense) by the square roots of $D_L(\mathbf{d})$ and $D_m(\mathbf{d})$. Both functions have zero value for $\mathbf{d} = \mathbf{0}$, and tend to increase, albeit not necessarily in a monotonic way, for increasing values of the distance $d = |\mathbf{d}|$. Note that, generally, the change in brightness does not only depend on $d$, but also on the displacement direction $\hat{\mathbf{d}} = \mathbf{d}/|\mathbf{d}|$, where the symbol ^ stands for "unit vector". The function $\sqrt{D_L(\mathbf{d})}$ provides the expected rms brightness change in absolute, energy-related, light units cd/m$^2$ or weighted Wm$^{-2}$sr$^{-1}$. $\sqrt{D_m(\mathbf{d})}$, in turn, gives us the relative change in mag$_V$/arcsec$^2$, that is, as $-2.5$ times the logarithm of the ratio of the final to the initial brightness.

Whether to use one or another function for estimating the maximum sampling distance depends on the way of specifying the desired reconstruction accuracy goals. $\sqrt{D_L(\mathbf{d})}$ is the function of choice when the tolerance threshold $\gamma$ (the maximum allowed rms difference between the measured and true brightness at the intermediante area between sampling points) is expressed in cd/m$^2$ or Wm$^{-2}$sr$^{-1}$.

$\sqrt{D_m(\mathbf{d})}$ shall be used, in turn, if the tolerance is expressed as a maximum allowed relative error, e.g. ±0.1 $mag_V/arcsec^2$. Once the goal is specified, the optimum sampling distance $d_s$ can be chosen as the value of $d$ at which the square root of the corresponding structure function attains the desired threshold. Note that $d_s$ will generally depend on the displacement direction $\hat{\mathbf{d}}$. A conservative, uniform, and relatively small value of $d_s$ may be adopted for all displacement directions, or a different value $d_s(\hat{\mathbf{d}})$ may be used for each direction if some information about the spatial distribution of the zenithal night sky brightness, or the artificial sources that produce it, is available a priori.

### B. The spatial spectrum of the zenithal sky brightness and the Nyquist-Shannon sampling theorem

The classical version of the Nyquist-Shannon sampling theorem (Papoulis 1981) states that a spectrally band-limited time signal $f(t)$ can be *exactly* reconstructed from a discrete set of samples, $f(t_n)$, taken at periodic times separated by a fixed interval $\tau_s$ (that is, at $t_n = t_o + n\tau_s$, with arbitrary $t_o$, and $n$ integer), as far as $\tau_s \leq 1/(2\nu_s)$, where $\nu_s$ is the maximum temporal frequency (Hz) present in the signal spectrum. The limiting time interval $\tau_s = 1/(2\nu_s)$ is known as the Nyquist sampling rate. The reconstruction of the original signal from this discrete set of samples is carried out by a low-pass filtering of the spectrum of the sampled signal, followed by an inverse Fourier transform to go back to the time domain. The net result of these operations is equivalent to interpolating the signal between samples by using a set of scaled $\mathrm{sinc}_n(t) = \sin[2\pi\nu_s(t - t_n)]/[2\pi\nu_s(t - t_n)]$ functions centered at the sampling points (Papoulis 1981). The most interesting feature of this theorem is that this interpolation does not merely approximate the values of the function between measurement points, but provides an *exact* reconstruction of $f(t)$ (noise propagation aside) for *all values* of $t$.

The Nyquist-Shannon theorem in the one-dimensional time domain can be easily extended to spaces of higher dimension. A classical two-dimensional formulation was developed fifty years ago for applications in the field of coherent optics (Papoulis 1981,

Goodman 1996). Adapted to our present issue, let us define the two-dimensional spectrum $\Lambda(\mathbf{v})$ of $L(\mathbf{r})$ as the Fourier transform:

$$\Lambda(\mathbf{v}) = \iint_{\infty} L(\mathbf{r}) \exp(-i2\pi\mathbf{v} \cdot \mathbf{r}) \, d^2\mathbf{r}, \tag{7}$$

where $\mathbf{v} = (v_x, v_y)$ is a vector whose components (units m$^{-1}$) play a role analogous to the time frequency, but now along the two orthogonal dimensions of the inverse space domain. The Nyquist-Shannon theorem states that if the function $L(\mathbf{r})$ is band-limited, that is, if its spectrum $\Lambda(\mathbf{v})$ becomes zero for spatial frequencies $v_x > v_{x\max}$ and $v_y > v_{y\max}$, then $L(\mathbf{r})$ can be exactly reconstructed at all points of its definition domain from a discrete set of samples $L(\mathbf{r}_{pq})$ taken at a rectangular grid of points $\mathbf{r}_{pq} = (x_o + pd_{sx}, y_o + qd_{sy})$, with arbitrary $(x_o, y_o)$, $p$ and $q$ integers, as far as $(d_{sx}, d_{sy}) \leq (1/(2v_{x\max}), 1/(2v_{y\max}))$. The latter condition states that the spatial sampling period along each orthogonal direction shall be smaller than half the inverse of the maximum spatial frequency present in the signal spectrum along that direction. The exact reconstruction of the original function $L(\mathbf{r})$ is carried out, analogously to the one-dimensional time case, by a low pass filtering of the spectrum of $L(\mathbf{r}_{pq})$ in the spatial frequency domain, followed by an inverse two-dimensional Fourier transform to go back to the space domain. This is equivalent to using a set of scaled two-dimensional sinc functions, centered at the sampling points, to exactly interpolate the values of $L(\mathbf{r})$ between samples. For an optimum sampling scheme, i.e. sampling at the Nyquist rate $(d_{sx}, d_{sy}) = (1/(2v_{x\max}), 1/(2v_{y\max}))$, the interpolating functions have the form (Goodman 1996):

$$\mathrm{sinc}_{pq}(x, y) = \frac{\sin[2\pi v_{x\max}(x - x_p)]\sin[2\pi v_{y\max}(y - y_q)]}{[4\pi^2 v_{x\max} v_{y\max}(x - x_p)(y - y_q)]}, \tag{8}$$

Analogous expressions and results can be obtained for the Fourier transform pair formed by $m(\mathbf{r})$ and its spatial spectrum, $M(\mathbf{v})$, related by Eq. (7) after the appropriate substitution of symbols.

## 3. Results

### 3.1. Artificial zenithal NSB

In order to get some insight about the optimal sampling distance in areas with different artificial light source distributions, we selected four 720x720 km$^2$ regions of the world and read the predicted values of their artificial zenithal night sky brightness from the NWA floating point dataset (Falchi et al. 2016, 2016b). Fig. 1 shows the corresponding brightness maps, displayed in mag$_V$/arcsec$^2$ units, to help visualizing the large dynamic range of the signal. From left to right and top to bottom, the selected regions are centered in Santiago de Compostela (Galicia, Spain), Berlin (Germany), the Joshua Tree National Park (USA), and Swan Hill (Australia). These areas have different mixes of highly populated cities, rural nuclei, extended oceanic waters and relatively unpopulated lands. As it would be anticipated, their average zenithal night sky brightnesses are very different, depending on the intensity and spatial distribution of the artificial light sources. The average values of the artificial component of the zenithal NSB in the central 74x74 km$^2$ part of these regions, separately computed in absolute and relative units, are: Santiago de Compostela (0.29 mcd/m$^2$, 21.7 mag$_V$/arcsec$^2$), Berlin (0.61 mcd/m$^2$, 21.1 mag$_V$/arcsec$^2$), Joshua Tree Nat. Park (0.17 mcd/m$^2$, 23.5 mag$_V$/arcsec$^2$), and Swan Hill (0.004 mcd/m$^2$, 27.6 mag$_V$/arcsec$^2$).

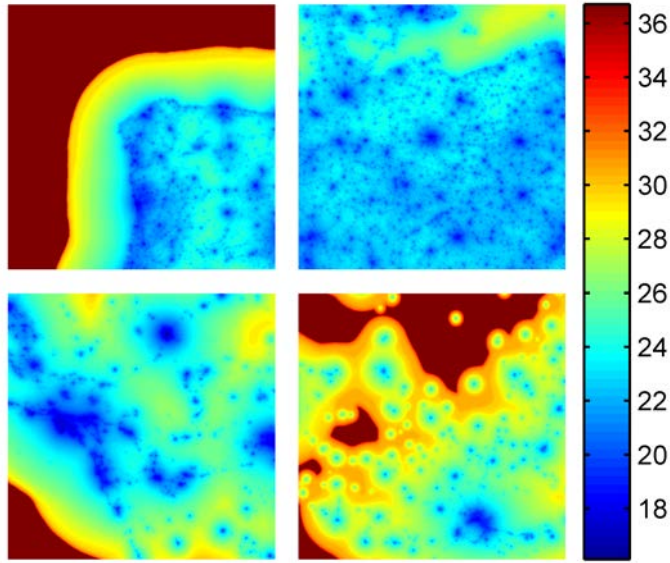

**Figure 1:** Artificial zenithal night sky brightness $m(\mathbf{r})$ in four regions of the world, according to the NWA estimates (Falchi et al. 2016, 2016b). Colour scale in mag$_V$/arcsec$^2$. Each region is about 720x720 km$^2$, and their centers are located at (a) Upper left: Santiago de Compostela, Galicia, Spain (42.8° N, 8.5° W); (b) Upper right: Berlin, Germany (52.5° N, 13.4° E); (c) Lower left: Joshua Tree National Park, California, USA (33.9° N, 115.9° W); (d) Lower right: Swan Hill, Australia (35.5° S, 143.6° E).

Fig. 2 displays the values of $\sqrt{D_L(\mathbf{d})}$ versus the displacement $d$, along the orthogonal latitude and longitude axes, for the central 74x74 km$^2$ area of the four regions depicted in Fig. 1. Positive values of $d$ correspond to displacements towards the South (blue), and East (red), respectively. As expected, the absolute brightness differences for a given value of the displacement are widely different from one region to another. Note also that the absolute value of the slope of $\sqrt{D_L(\mathbf{d})}$ tends to decrease with the absolute displacement. For very large displacements the structure function is expected to saturate: when $d$ is sufficiently large in comparison with the correlation length of the zenithal sky brightness, the correlation function $B_L(\mathbf{d})$ tends to the squared average of the brightness, $\langle L \rangle^2$, and hence, according to Eq. (5), the structure function tends to a constant value, equal to $2(\sigma_L{}^2 - \langle L \rangle^2)$.

Fig. 3 displays the expected rms change in brightness in relative $mag_V/arcsec^2$ units, $\sqrt{D_m(\mathbf{d})}$. The behaviour, in this case, is remarkably similar for all four sites: the slope of $\sqrt{D_m(\mathbf{d})}$ for small values of $d$ is approximately equal to 0.1 $mag_V/arcsec^2$ per km in each case. Of course, the fact that $\sqrt{D_m(\mathbf{d})}$ has some definite value does not mean that all pixels experience this precise amount of change. Fig. 4 shows the cumulative histogram of the changes in magnitude that a set of observers would record after traveling 1 pixel (0.927 km) southwards, evaluated in the central 74x74 $km^2$ area of the maps in Fig. 1. Each analyzed region presents a particular histogram signature, with about 69%-87% of adjacent pixels having an absolute brightness difference smaller than 0.1 $mag_V/arcsec^2$.

Figs. 5 and 6 show the spectral power densities $|\Lambda(\mathbf{v})|^2$ and $|M(\mathbf{v})|^2$ of the zenithal night sky brightness maps $L(\mathbf{r})$ and $m(\mathbf{r})$ expressed in absolute and relative units, respectively, taken along two orthogonal axes (longitude and latitude). The maximum spatial frequency contained in these maps is equal to one half the inverse spatial pixel size, that is $v_{y,\,max}$ =0.54 $km^{-1}$ along the latitude axis and $v_{x,\,max}$ =0.54/cos$\phi$ $km^{-1}$ along the longitude axis, where $\phi$ is the latitude. The inspection of these figures does not allow to identify a well-definite spatial cut-off frequency beyond which the spectra become identically zero. This suggests that the Nyquist sampling rate for the zenithal night sky brightness distribution will likely be higher than one sample per km.

Although the spatial spectra in Fig. 5 and 6 do not become strictly zero within the displayed frequency range, their relative values for spatial frequencies larger than ~0.25 $km^{-1}$ are quite small. This means that these high-frequency harmonic components contribute in a minor amount to the zenithal NSB, and it suggests that an approximate reconstruction of the zenithal NSB map of the regions under study could be obtained by sampling at points separated by ~2 km. Note that this period is larger than the expected Nyquist sampling rate (the signal is undersampled), so the Nyquist-Shannon theorem does not strictly apply and no exact reconstruction of the brightness between samples is to be expected. However, an approximate reconstruction may be useful for many practical purposes. As an example, Fig. 7 shows the approximate reconstruction of a zenithal sky brightness map from an undersampled signal. Fig. 7(a) shows the central 74x74 km$^2$ part of the region around Santiago de Compostela. Fig. 7(b) shows the brightness samples taken at about every 2 km along the orthogonal latitude/longitude axes. Due to the discrete pixel structure of the map, the actual sampling periods are 1.85 km (2 pixels) along the latitude (vertical) axis and 2.04 km (3 pixels) along the longitude (horizontal) one. Note that for the purposes of this calculation these samples are assumed to be taken not only in the restricted 74x74 km$^2$ area, but across the whole region displayed in Fig. 1 (top left). Fig. 7(c) shows the modulus of the Fourier spectrum of the sampled signal (whole region). This spectrum was subsequently multiplied in the spatial frequency domain by a two-dimensional super-Gaussian filter $G(\nu_x, \nu_y) = \exp\{-[(\nu_x/\nu_{cx})^8 + (\nu_y/\nu_{cy})^8]\}$, with $\nu_{cx}$ =0.25 km$^{-1}$ and $\nu_{cy}$ =0.27 km$^{-1}$, Fig. 7(d), and the result, Fig. 7(e), was transformed back to the spatial domain by an inverse Fourier transform, Fig. 7(f). The final result closely ressembles the actual brightness distribution although, under close inspection, it can be seen that the smallest features become somewhat blurred and some degree of detail is lost, due to undersampling. The root mean squared difference between the original and the reconstructed map in this central area is 0.0016 mag$_V$/arcsec$^2$, although individual pixels may show substantially larger differences, in the range (-0.33, 0.23) mag$_V$/arcsec$^2$.

Note that this reconstruction was carried out making use of the samples taken in a region substantially wider than the one shown in Fig. 7(b). Since the filtering procedure described above is equivalent to interpolating the samples with two-dimensional sinc functions centered at the sampling points, restricting the sampling domain to the small region shown in Fig. 7(b) would give rise to a highly inaccurate reconstruction at the rim of the image, for want of the required contributions from neighbouring points outside this border.

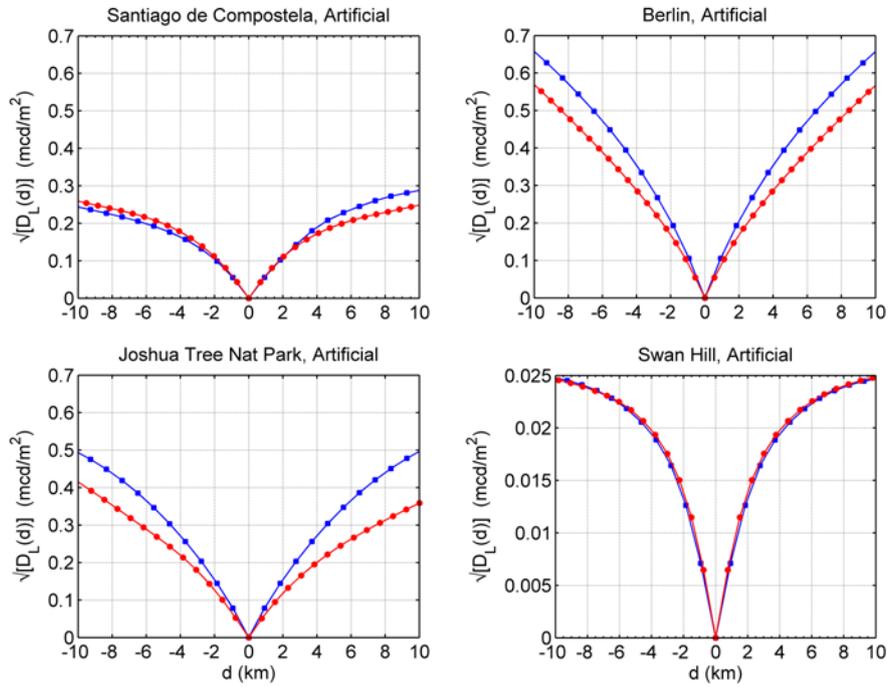

**Figure 2:** $\sqrt{D_L(\mathbf{d})}$, the square root of the artificial luminance structure function in mcd/m², versus the displacement *d* in km, evaluated in the central 74x74 km² area of the four regions depicted in Fig. 1. The function is shown for displacements along two orthogonal axes, latitude and longitude. Positive values of *d* correspond to displacements towards the South (blue), and East (red), respectively. Note that the expected rms change in absolute brightness for any given displacement strongly depends on the typical brightness levels of each region.

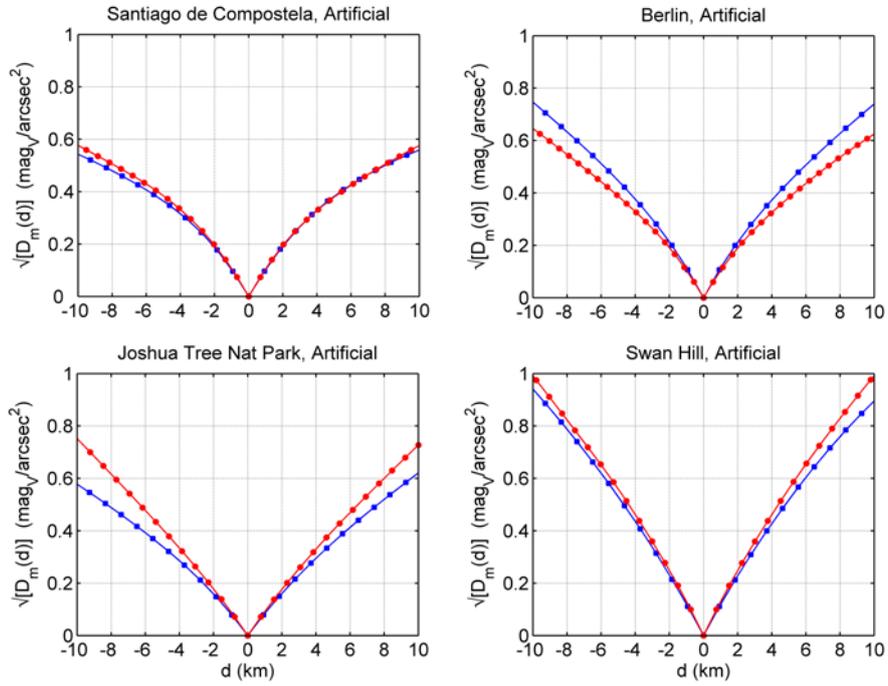

**Figure 3:** $\sqrt{D_m(\mathbf{d})}$, the square root of the artificial luminance structure function in mag$_V$/arcsec$^2$, versus the displacement $d$ in km, evaluated in the central 74x74 km$^2$ area of the four regions depicted in Fig. 1. Positive values of $d$ correspond to displacements towards the South (blue), and East (red), respectively. Note that the expected rms change in mag$_V$/arcsec$^2$ for small displacements is fairly similar in the four areas, irrespective of the absolute brightness of each one. An rms change of 0.1 mag$_V$/arcsec$^2$ is consistently achieved for displacements of order ~1 km in all four cases.

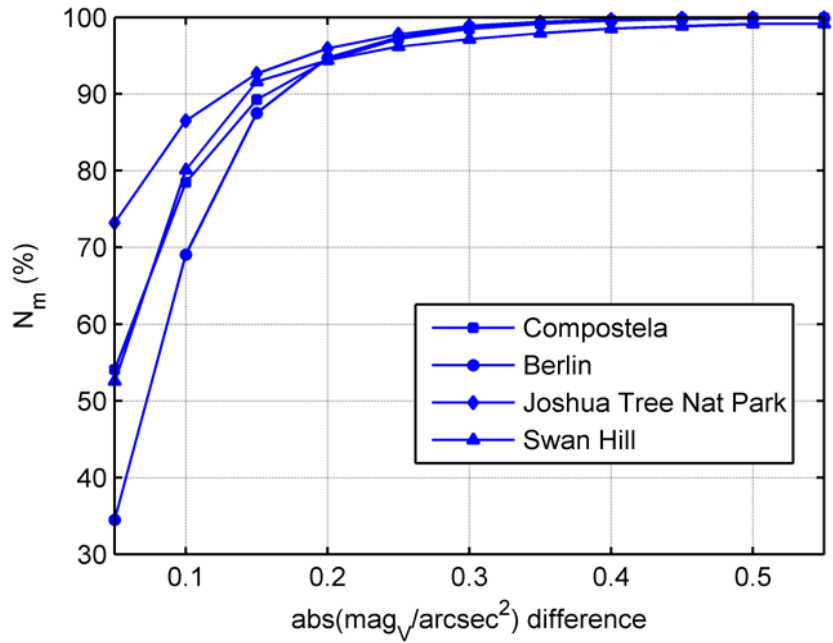

**Figure 4:** Cumulative histogram of the changes in artificial brightness after travelling one pixel southwards (0.927 km), evaluated in the central 74x74 km$^2$ area of the four regions depicted in Fig. 1. $N_m$ is the number of pixels (in %) that change their brightness by an absolute amount smaller than the value indicated in the horizontal axis, binned in 0.05 mag$_V$/arcsec$^2$ intervals.

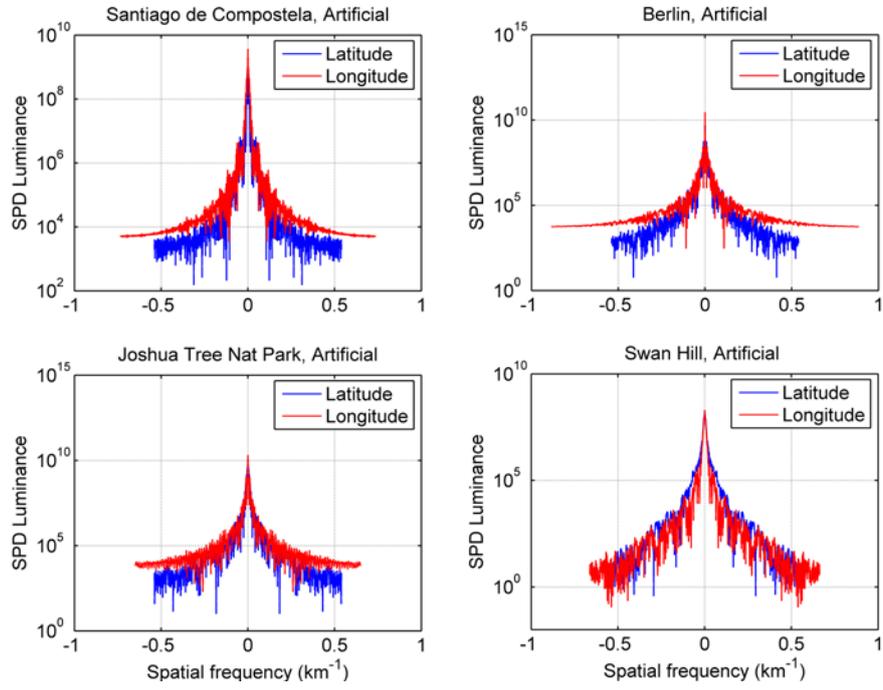

**Figure 5:** Spatial spectral power density $\left|\Lambda(\mathbf{v})\right|^2$ along two orthogonal directions (the local latitude and longitude axes) of the artificial sky brightness distribution $L(\mathbf{r})$ (mcd/m²) in the four regions displayed in Fig. 1.

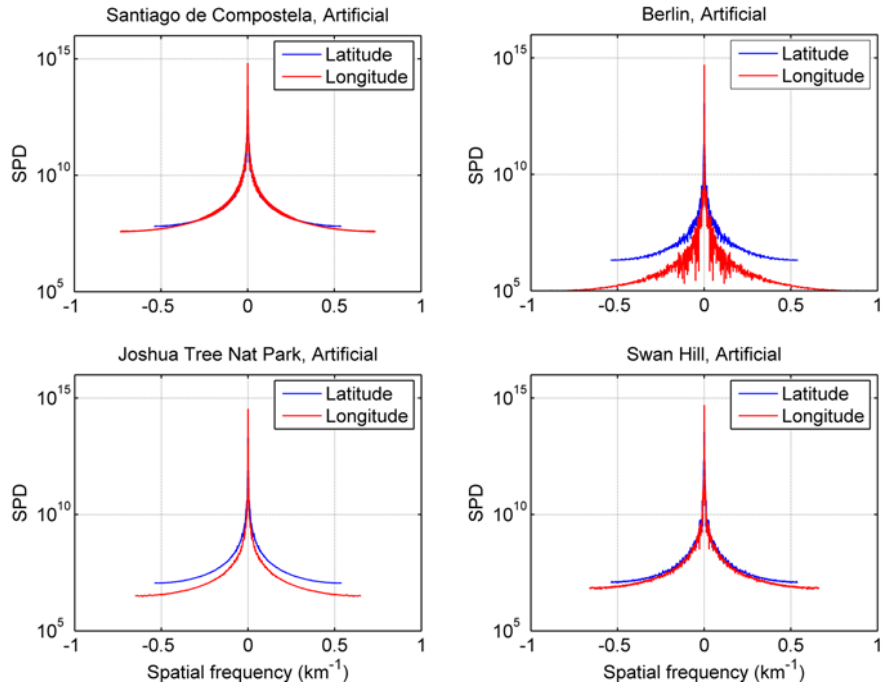

**Figure 6:** Spatial spectral power density $\left|\mathrm{M}(\mathbf{v})\right|^2$ along two orthogonal directions (the local latitude and longitude axes) of the artificial sky brightness distribution $m(\mathbf{r})$ (mag$_V$/arcsec$^2$) in the four regions displayed in Fig. 1.

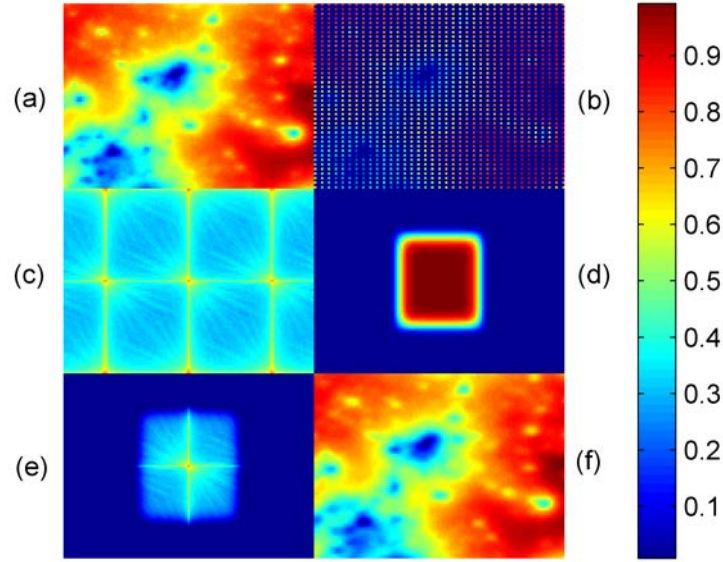

**Figure 7:** Sampling and reconstruction of the continuous artificial zenithal NSB distribution $m(\mathbf{r})$, in the 720x720 km² region centered at Santiago de Compostela (42.8° N, 8.5° W), using the Nyquist-Shannon theorem. (a) Artificial zenithal night sky brightness $m(\mathbf{r})$ computed from the luminances of the NWA floating point dataset; (b) Samples of the previous map taken about every 2 km in the North-South and East-West directions (see text for details); (c) Modulus of the two-dimensional spectrum (Fourier transform) of the sampled map; (d) Two-dimensional rectangular low-pass filter with super-Gaussian profile ($n$=8) in the spatial frequency domain; (e) Modulus of the low-pass filtered spectrum of the sampled map; (f) inverse Fourier transform of the low-pass filtered spectrum. The zenithal night sky brightness distribution reconstructed from the discrete set of samples closely resembles the original map. To facilitate the visualization of the middle and high spatial frequency regions of the spectrum, the images (c) and (e) are displayed in a logarithmic scale. In order to discern the individual sampling points, in images (a), (b) and (f), only the central 74x74 km² of the whole region are shown. The colourbar is scaled to 1 for the maximum and 0 for the minimum value of each individual image.

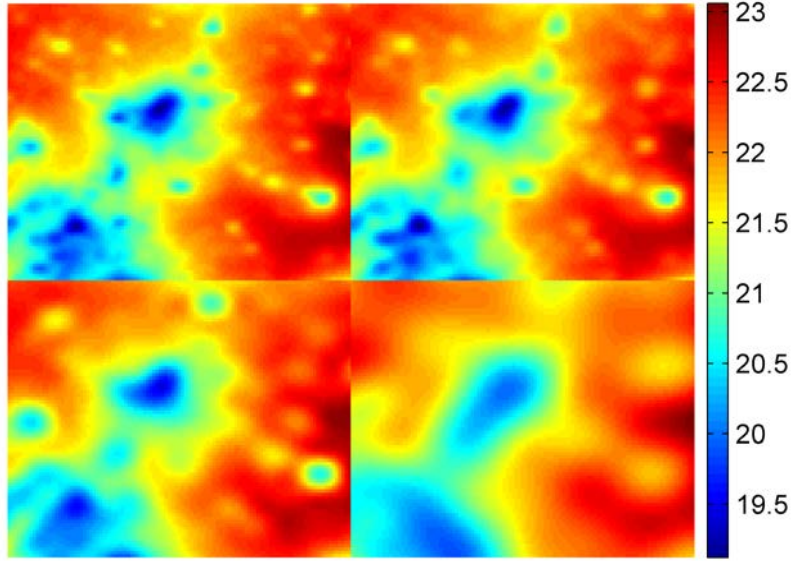

**Figure 8:** Reconstructed artificial zenithal NSB distributions, $m(\mathbf{r})$, using different undersampling periods. Upper left: 2 km, the same as in Fig. 7 (f); upper right: 3 km; lower left: 5 km; lower right: 10 km. The area shown in the images corresponds to the central 74x74 km$^2$ region around Santiago de Compostela (42.8° N, 8.5° W). Scale in mag$_v$/arcsec$^2$.

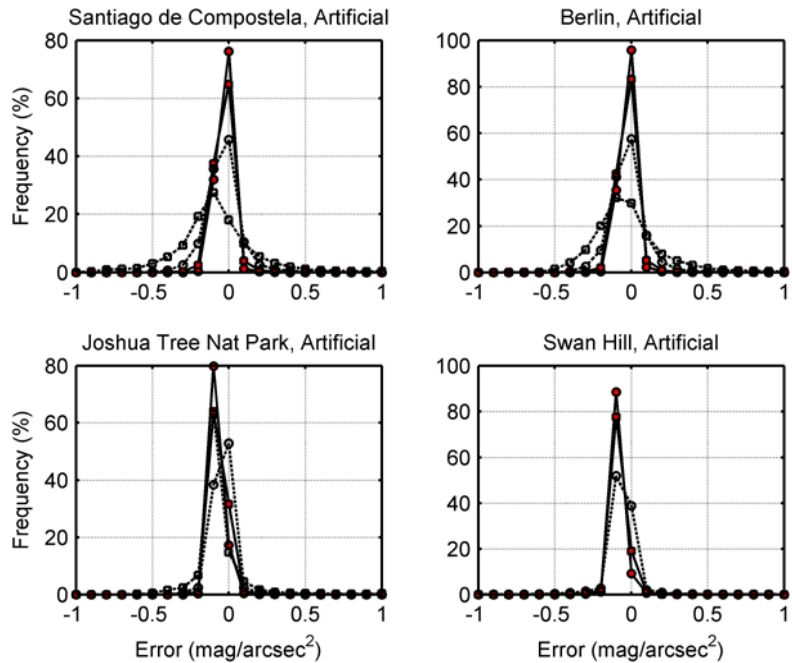

**Figure 9:** Histograms of the zenithal NSB reconstruction errors (in mag$_v$/arcsec$^2$) due to undersampling for the four locations analyzed in this work. Sampling periods: solid circles, 2 km; solid squares, 3 km; open circles, 5 km; open squares, 10 km.

If the sampling is coarser the accuracy in the reconstruction is expected to decrease. Figure 8 displays the retrieved artificial zenithal NSB distributions in the 74x74 km$^2$ area surrounding Santiago de Compostela for sampling periods 2 km, 3 km, 5 km, and 10 km, respectively, with minor differences along each axis due to the different pixel size. It can be seen that undersampling with a period of 3 km still preserves many details of the original distribution shown in Fig 7 (a). Undersampling with larger periods, such as 5 and 10 km, gives rise to a noticeably blurring of the retrieved features. The error histograms of the reconstructed maps for the whole 720x720 km$^2$ regions around each observation point are shown in Fig. 9

### 3.2. Total zenithal NSB

The total zenithal NSB includes not only the artificial brightness but also the contribution of the natural sources, most notably the celestial objects located above the observer, expressed in energy or light units (Wm$^{-2}$sr$^{-1}$ or cd/m$^2$). The total zenithal NSB is a directly measurable physical quantity, and can also be quantitatively estimated by adding to the artificial NSB component the brightness of the natural sky computed by means of models such as the one developed by Duriscoe (2013). For the purposes of this work we will assume a constant natural sky contribution of order 0.174 mcd/m$^2$, equivalent to 22 mag$_V$/arcsec$^2$ (Falchi et al. 2016). Figure 10 displays the total zenithal NSB maps of the regions under study, in logarithmic scale.

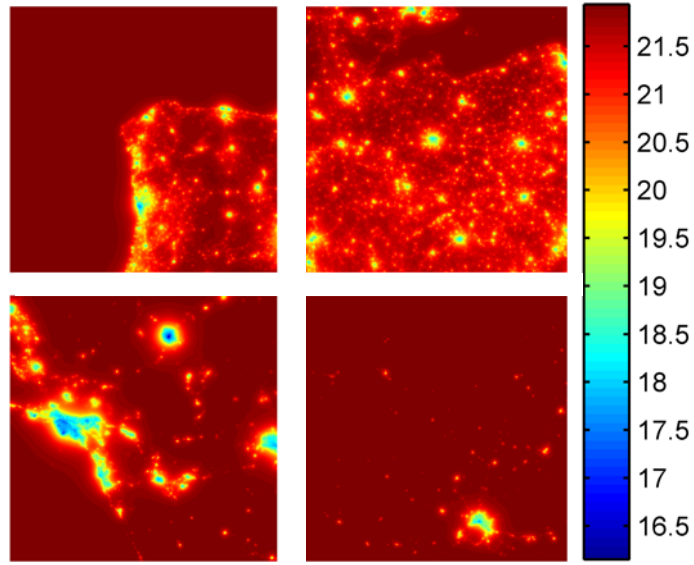

**Figure 10:** Total zenithal night sky brightness $m(\mathbf{r})$ in the four regions of the world analyzed in this work, according to the NWA estimates (Falchi et al. 2016, 2016b) with an assumed natural sky contribution of 0.174 mcd/m$^2$ (equivalent to 22 mag$_V$/arcsec$^2$). Colour scale in mag$_V$/arcsec$^2$. Each region is about 720x720 km$^2$ wide, and their centers are located at (a) Upper left: Santiago de Compostela, Galicia, Spain (42.8° N, 8.5° W; (b) Upper right: Berlin, Germany (52.5° N, 13.4° E); (c) Lower left: Joshua Tree National Park, California, USA (33.9° N, 115.9° W); (d) Lower right: Swan Hill, Australia (35.5° S, 143.6° E).

Since the natural sky brightness simply adds a constant value in energy or light units to the artificial one, the $\sqrt{D_L(\mathbf{d})}$ function remains invariant, and the results presented in the previous section directly hold for the total zenithal NSB. Essentially the same happens to the Fourier spectrum $\Lambda(\mathbf{v})$ of the total brighness, whose only modification is an increase of its value at the origin, corresponding to the null spatial frequency $\mathbf{v} = (0,0)$, which reflects the average value of the zenithal NSB across the geographical region considered. However, if the brightness is expressed in logarithmic mag$_V$/arcsec$^2$ units this invariance does not hold, due to the non-linearity of that brightness scale. This can be seen in the behaviour of the total $\sqrt{D_m(\mathbf{d})}$ function, shown in Fig. 11, as well as in the cumulative histogram corresponding to a displacement of one pixel southwards, shown in Fig 12. The total magnitude change per km of displacement in direction South is no longer equal in all four regions. As a matter of fact, the existence of a minimum level of zenithal NSB, due to the natural sky contribution, forces the slope of the $\sqrt{D_m(\mathbf{d})}$ structure function for the total brighness to be smaller than the one for the artificial component alone. This effect is of minor importance in densely populated regions where the artificial contribution is much higher than the natural one (e.g. Santiago de Compostela or Berlin, which show a behaviour not unlike the one displayed in Fig. 3, although with smaller values) but clearly affects those areas with low levels of artificial light (e.g. Swan Hill).

The slices of the spectral power distribution $|M(\mathbf{v})|^2$ of the total zenithal NSB in the four areas under study are displayed in Fig. 13. Again, no clear cut-off frequency is apparent from the plots, although the fast decay of the SPD enables an approximate reconstruction using a moderate degree of undersamplig, as shown above for the artificial zenithal NSB case. Fig. 14 illustrates this possibility, with samples spaced every 2 km in the Santiago de Compostela region (see Section 3.1. for details).

Larger sampling periods will give rise to additional amounts of blurring in the estimated total zenithal NSB maps. Figure 15 shows the reconstructed brightness in the 74x74 km$^2$ area surrounding Santiago de Compostela for sampling periods 2 km, 3 km, 5 km, and 10 km, respectively. As in the artificial NSB case, undersampling with a 3 km period still provides a reasonably good reconstruction of the original distribution, shown in Fig 14 (a). The performance is degraded for undersampling of 5 and 10 km. The error histograms of the reconstructed maps for the whole 720x720 km$^2$ regions around each observation point are shown in Fig. 16.

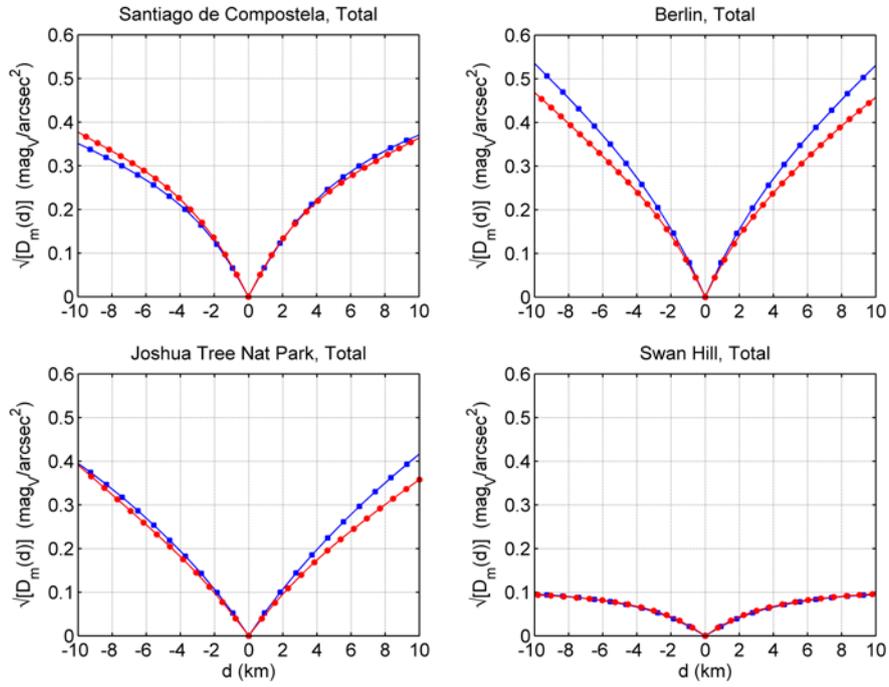

**Figure 11:** $\sqrt{D_m(\mathbf{d})}$, the square root of the total luminance structure function in mag$_V$/arcsec$^2$, versus the displacement $d$ in km, evaluated in the central 74x74 km$^2$ area of the four regions depicted in Fig. 1. Positive values of $d$ correspond to displacements towards the South (blue), and East (red), respectively. Note that the expected rms change in mag$_V$/arcsec$^2$ for small values of $d$ is no longer equal in all four areas, and is smaller than the one for the artificial component (Fig. 3).

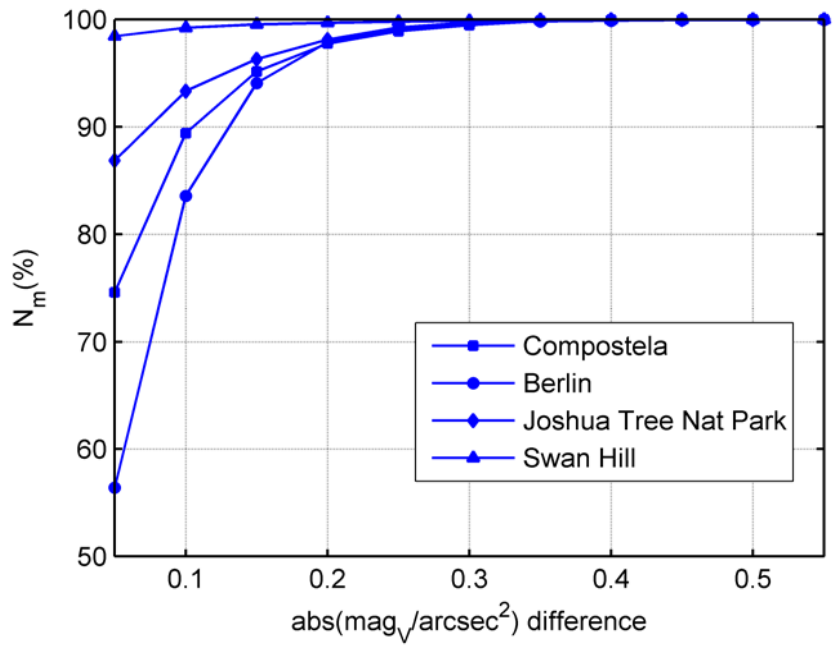

**Figure 12:** Cumulative histograms of the changes in the total brighness after travelling one pixel southwards (0.927 km), evaluated in the central 74x74 $km^2$ area of the four regions depicted in Fig. 8. $N_m$ is the number of pixels (in %) that change their brightness by an absolute amount smaller than the value indicated in the horizontal axis, with 0.05 $mag_v/arcsec^2$ bins.

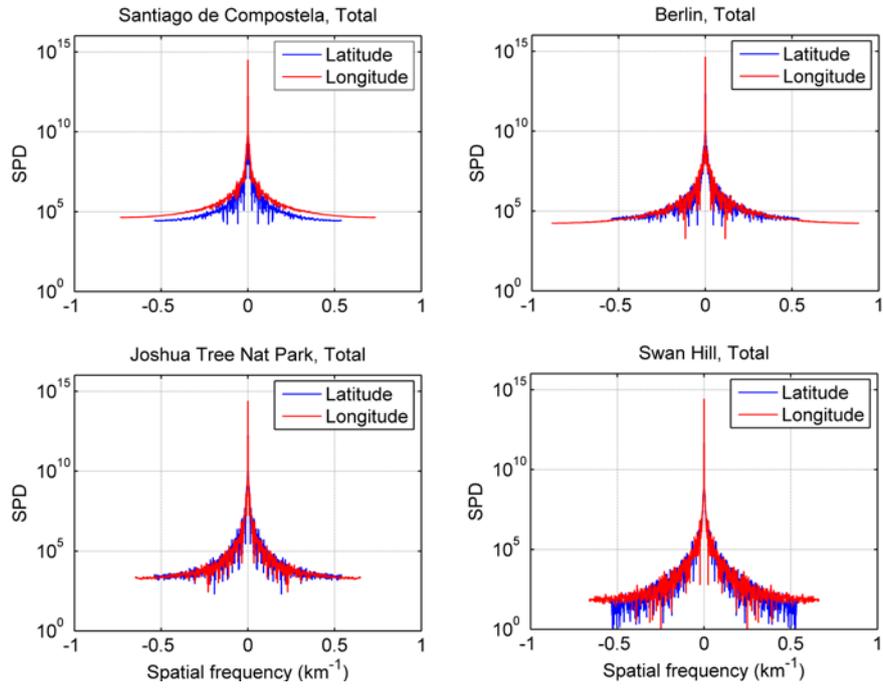

**Figure 13:** Spatial spectral power density $\left|M(\mathbf{v})\right|^2$ along two orthogonal directions (the local latitude and longitude axes) of the total zenithal sky brightness distribution $m(\mathbf{r})$ (mag$_V$/arcsec$^2$) in the four regions displayed in Fig. 8.

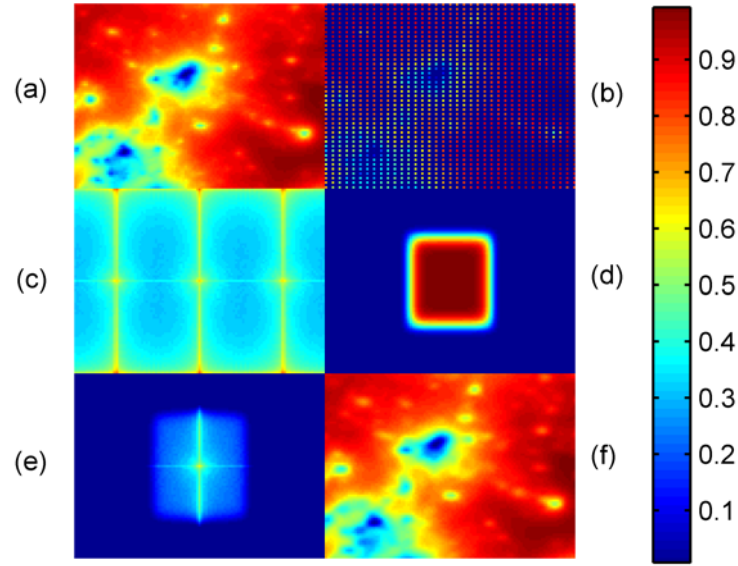

**Figure 14:** Sampling and reconstruction of the continuous total zenithal night sky brightness distribution, $m(\mathbf{r})$, in the 720x720 km$^2$ region centered at Santiago de Compostela (42.8° N, 8.5° W) using the Nyquist-Shannon theorem. (a) Total zenithal night sky brightness $m(\mathbf{r})$ computed from the luminances of the NWA floating point dataset, adding the natural sky contribution; (b) Samples of the previous map taken every 2 km in the North-South and East-West directions; (c) Modulus of the two-dimensional spectrum (Fourier transform) of the sampled map; (d) Two-dimensional rectangular low-pass filter with super-Gaussian profile ($n$=8) in the spatial frequency domain; (e) Modulus of the low-pass filtered spectrum of the sampled map; (f) inverse Fourier transform of the low-pass filtered spectrum. The zenithal night sky brightness distribution reconstructed from the discrete set of samples closely resembles the original map. To facilitate the visualization of the middle and high spatial frequency regions of the spectrum, the images (c) and (e) are displayed in a logarithmic scale. In order to discern the individual sampling points in images (a), (b) and (f), only the central 74x74 km$^2$ of the region are shown. The colourbar is scaled to 1 for the maximum and 0 for the minimum value of each individual image.

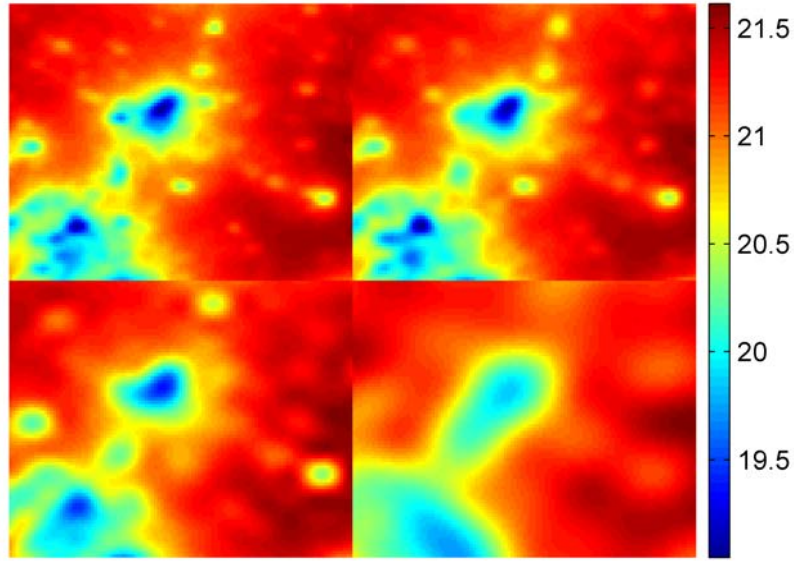

**Figure 15:** Reconstructed total zenithal night sky brightness distributions, $m(\mathbf{r})$, for different undersampling periods. Upper left: 2 km, the same as in Fig. 14 (f); upper right: 3 km; lower left: 5 km; lower right: 10 km. The area shown corresponds to the central 74x74 km² region around Santiago de Compostela (42.8° N, 8.5° W). Scale in $mag_V/arcsec^2$.

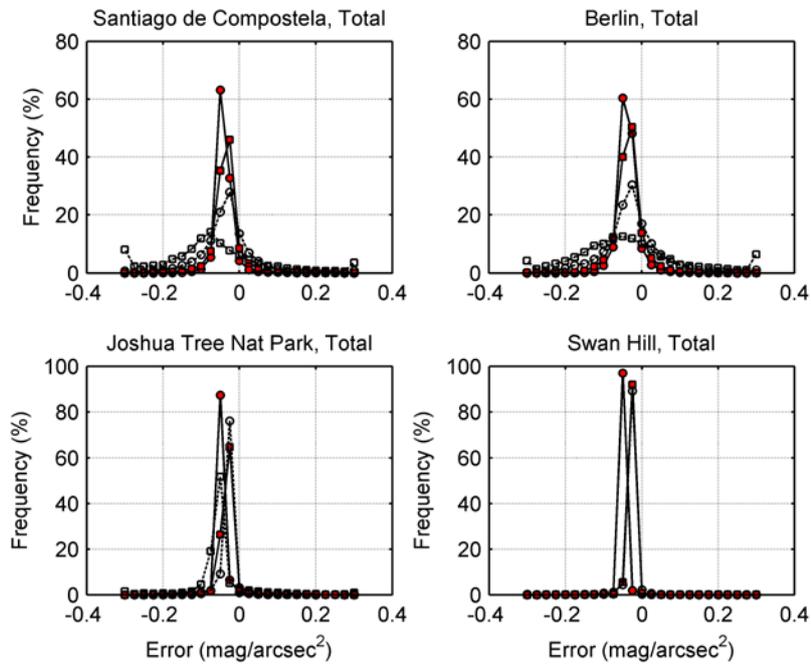

**Figure 16:** Histograms of the reconstruction errors due to undersampling for the four locations analyzed in this work. Sampling periods: Solid circles, 2 km; solid squares, 3 km; open circles, 5 km; open squares, 10 km.

**4. Discussion**

The results presented in this work are intended to be a first approximation to the problem of determining the optimum sampling distance to retrieve the continuous zenithal night sky brightness in a region. Two approaches have been explored. The first, based on the behaviour of the spatial structure function, provides what can be called a *weak reconstruction criterion*, that is, a sampling distance that guarantees that, in the rms sense, the change in brightness between consecutive sampling points will not exceed a given predefined value. According to this criterion, about one measurement per square kilometer could be sufficient for determining the artificial zenithal night sky brightness at any point of a region to within 0.1 mag$_V$/arcsec$^2$ (in the root-mean-square sense) of its true value in the Johnson-Cousins V band.

The second approach, based on the Nyquist-Shannon spatial sampling theorem, provides a *strong reconstruction criterion*, in the sense that sampling at the Nyquist rate guarantees that an exact reconstruction of the true zenithal night sky brightness distribution can be obtained for all points of the region. Based on the presently available datasets, the precise value of the required Nyquist rate is still unclear, but will probably be more than one sample per km. An approximate reconstruction of the original brightness can nevertheless be obtained by means of a moderate undersampling: the scenarios described in Section 3 using one sample every two or three km show the expected performance. Note however that reconstructing a two-dimensional function using the Nyquist-Shannon approach within a finite domain often requires acquiring a huge number of samples in the region surrounding it, which can be impractical in many situations.

Other reconstruction schemes, in particular least-squares fits of the zenithal NSB by different families of orthogonal polynomials whose domain of definition be coincident with the region of interest, could prove to be more efficient for retrieving the continuous zenithal brightness maps from finite sets of samples. Kriging is also a useful technique for estimating continuous spatial distributions from a discrete set of measurements (Fersch & Walker, 2012). These approaches, which do not require using uniformly spaced sampling grids, have very interesting features for dealing with this problem and deserve further research.

Note that the results presented in this paper address the basic question of the sampling and reconstruction of the zenithal night sky brightness in extended territories under a layered atmosphere with constant atmospheric conditions, as assumed in the calculation of *The new world atlas of artificial night sky brightness* (Falchi et al., 2016). The structure functions and the spatial Fourier spectra of the actual zenithal night sky brightness distributions have additional degrees of variability, due to the unavoidable spatial and temporal inhomogeneities in the aerosol concentration profiles, as well as to the changing course of the anthropogenic emissions of light and the natural brightness contributed by the celestial bodies located above the observer. Estimating the artificial component of the zenithal sky brightness requires subtracting the natural component from the measurements, and the resulting values shall be corrected for the atmospheric conditions and for the variable time course of the artificial light emissions throughout the night.

As a final remark, the methods here described can be equally applied to the reconstruction of the night sky brightness in any arbitrary direction of the celestial hemisphere (not just the zenith), from a discrete set of samples taken in different points of the region. The optimum sampling distance does not have to be the same for all hemispheric directions, and the minimum of its values should be taken as a reference for carrying out theoretical calculations or planning observational field campaigns. As previously shown, the all-sky brightness distribution at any given site can be reconstructed from a finite number of samples taken in different directions of the celestial hemisphere (Bará et al. 2014, 2015, 2015b). An optimum sampling scheme, combining directional (across the upper hemisphere in the observer's reference frame) and spatial (across the region) measurements may open the way to the reconstruction of the all-sky night brightness distribution at any place in the region of interest at a lower computational cost.

## 5. Conclusions

Two different tools were used to determine the optimum sampling distance in order to retrieve, with sufficient accuracy, the continuous zenithal night sky brightness map across a wide region: the luminance structure function and the Nyquist-Shannon spatial sampling theorem. The analysis of sky brightness data for several regions of the world suggests that, as a rule of thumb, about one measurement per square kilometer could be sufficient for determining the artificial zenithal night sky brightness at any point of a region to within $\pm0.1$ $mag_V$/arcsec$^2$ (in the root-mean-square sense) of its true value in the Johnson-Cousins V band. The exact reconstruction of the zenithal night sky brightness map from samples taken at the Nyquist rate seems to be considerably more demanding.

## Acknowledgments

This work was developed within the framework of the Spanish Network for Light Pollution Studies (AYA2015-71542-REDT). Special thanks are due to James Irish for useful suggestions and comments. The availability of the floating point dataset of the *New world atlas of artificial night sky brightness* is gratefully acknowledged.

## References

Aceituno J., Sánchez S. F., Aceituno F. J., Galadí-Enríquez D., Negro J. J. , Soriguer R. C., and Sanchez-Gomez G., 2011, An all-sky transmission monitor: ASTMON. Publ. Astron. Soc. Pac., 123, 1076–1086

Aubé M., 2015, Physical behaviour of anthropogenic light propagation into the nocturnal environment. Phil. Trans. R. Soc. B 370:20140117. doi: 10.1098/rstb.2014.0117

Aubé M., Kocifaj M., Zamorano J., Solano Lamphar H.A., Sánchez de Miguel A., 2016, The spectral amplification effect of clouds to the night sky radiance in Madrid. Journal of Quantitative Spectroscopy & Radiative Transfer 181, 11–23

Bará S., Nievas M., Sánchez de Miguel A., Zamorano J., 2014, Zernike analysis of all-sky night brightness maps. Applied Optics, 53(12), 2677-2686. doi: 10.1364/AO.53.002677

Bará S., Tilve V., Nievas M., Sánchez de Miguel A., Zamorano J., 2015, Zernike power spectra of clear and cloudy light-polluted urban night skies. Applied Optics, 54(13), 4120-4129. doi: 10.1364/AO.54.004120

Bará S., Ribas S., Kocifaj M., 2015, Modal evaluation of the anthropogenic night sky brightness at arbitrary distances from a light source. Journal of Optics (IOP) 17, 105607. doi:10.1088/2040-8978/17/10/105607

Bará S., 2016, Anthropogenic disruption of the night sky darkness in urban and rural areas, Royal Society Open Science 3: 160541. doi: 10.1098/rsos.160541. Published 19 October 2016

Bessell M. S., 1979, UBVRI photometry II: The Cousins VRI system, its temperature and absolute flux calibration, and relevance for two-dimensional photometry. Publ. Astron. Soc. Pac. 91, 589–607

CIE Commision Internationale de l'Éclairage, 1990, *CIE 1988 2° Spectral Luminous Efficiency Function for Photopic Vision*. Vienna: Bureau Central de la CIE

Cinzano P., Elvidge C. D., 2004, Night sky brightness at sites from DMSP-OLS satellite measurements. Mon. Not. R. Astron. Soc. 353, 1107–1116 doi:10.1111/j.1365-2966.2004.08132.x

Cinzano P., Falchi F., 2012, The propagation of light pollution in the atmosphere. Mon. Not. R. Astron. Soc. 427, 3337–3357. doi:10.1111/j.1365-2966.2012.21884.x

Cinzano P., Falchi F., Elvidge, C., 2001, The first world atlas of the artificial night sky brightness. Mon. Not. R. Astron. Soc., 328, 689–707

Cinzano P., Falchi F., Elvidge C. D., 2001, Naked-eye star visibility and limiting magnitude mapped from DMSP-OLS satellite data. Mon. Not. R. Astron. Soc. 323, 34–46

den Outer P., Lolkema D., Haaima M., van der Hoff R., Spoelstra H., Schmidt W., 2015, Stability of the Nine Sky Quality Meters in the Dutch Night Sky Brightness Monitoring Network. Sensors, 15, 9466-9480. doi: 10.3390/s150409466

Duriscoe D. M., 2013, Measuring Anthropogenic Sky Glow Using a Natural Sky Brightness Model, Publications of the Astronomical Society of the Pacific, 125(933), 1370-1382. http://www.jstor.org/stable/10.1086/673888

Duriscoe D., 2016, Photometric indicators of visual night sky quality derived from all-sky brightness maps. Journal of Quantitative Spectroscopy & Radiative Transfer, 181, 33–45. doi: 10.1016/j.jqsrt.2016.02.022

Duriscoe D. M., Luginbuhl C. B., Moore C. A., 2007, Measuring Night-Sky Brightness with a Wide-Field CCD Camera. Publications of the Astronomical Society of the Pacific, 119(852), 192-213

Espey B., McCauley J., 2014, Initial Irish light pollution measurements and a new Sky Quality Meter-based data logger. Lighting Res. Technol. 46, 67–77

Falchi F., 2011, Campaign of sky brightness and extinction measurements using a portable CCD camera. Mon. Not. R. Astron. Soc., 412, 33–48. doi: 10.1111/j.1365-2966.2010.17845.x

Falchi F., Cinzano P., Duriscoe D., Kyba C. C. M., Elvidge C. D., Baugh K., Portnov B. A, Rybnikova N. A., Furgoni R., 2016, The new world atlas of artificial night sky brightness. Sci. Adv. 2, e1600377. (doi: 10.1126/sciadv.1600377)

Falchi F., Cinzano P., Duriscoe D., Kyba C. C. M., Elvidge C. D., Baugh K., Portnov B., Rybnikova N. A., Furgoni, R., 2016b, Supplement to: The New World Atlas of

Artificial Night Sky Brightness. GFZ Data Services. http://doi.org/10.5880/GFZ.1.4.2016.001

Fersch A., Walker C., 2012, Light Pollution Around Tucson, AZ And Its Effect On The Spatial Distribution Of Lesser Long-nosed Bats. American Astronomical Society, AAS Meeting #219, id.141.04.  adsabs.harvard.edu/abs/2012AAS...21914104F

Garstang R. H., 1986, Model for artificial night-sky illumination. Publ. Astron. Soc. Pac. 98, 364-375

Gaston K. J., Bennie J., Davies T. W., Hopkins J., 2013, The ecological impacts of nighttime light pollution: a mechanistic appraisal. Biological Reviews 88, 912–927

Gaston K. J., Duffy J. P., Gaston S., Bennie J., Davies T. W., 2014, Human alteration of natural light cycles: causes and ecological consequences. Oecologia 176, 917–931

Ges X., Bará S., García-Gil M., Zamorano J., Ribas S. J., Masana E., 2017, Light pollution offshore: zenithal sky glow measurements in the Mediterranean coastal waters. https://arxiv.org/abs/1705.02508

Goodman J. W., 1996, *Introduction to Fourier Optics*, 2nd. ed. McGraw-Hill, New York. pp. 22-26

Hölker F., Wolter C., Perkin E. K., Tockner K., 2010, Light pollution as a biodiversity threat. Trends in Ecology and Evolution 25, 681-682

Hölker F., Moss T., Griefahn B., Kloas W., Voigt C. C., Henckel D., Hänel A., Kappeler P. M., Völker S., Schwope A., Franke S., Uhrlandt D., Fischer J., Klenke. R., Wolter C., Tockner K., 2010b, The dark side of light: a transdisciplinary research agenda for light pollution policy. Ecology and Society 1 (4): 13. (www.ecologyandsociety.org/vol15/iss4/art13/)

Jechow A., Kolláth Z., Lerner A., Hölker F., Hänel A., Shashar N., Kyba C. C. M., 2017, Measuring Light Pollution with Fisheye Lens Imagery from A Moving Boat, A Proof of Concept. arXiv:1703.08484v1 [q-bio.OT] 22 Mar 2017

Kocifaj M., 2007, Light-pollution model for cloudy and cloudless night skies with ground-based light sources. Applied Optics 46, 3013-3022

Kocifaj M., 2016, A review of the theoretical and numerical approaches to modeling skyglow: Iterative approach to RTE, MSOS, and two-stream approximation. Journal of Quantitative Spectroscopy & Radiative Transfer 181, 2–10

Kolláth Z., 2010, Measuring and modelling light pollution at the Zselic Starry Sky Park. Journal of Physics: Conference Series, 218, 012001. doi: 10.1088/1742-6596/218/1/012001

Kyba C. C. M. et al., 2015, Worldwide variations in artificial skyglow. Sci. Rep. 5, 8409. doi:10.1038/srep08409

Kyba C. C. M., Ruhtz T., Fischer J., Hölker, F., 2012, Red is the new black: how the colour of urban skyglow varies with cloud cover. Mon. Not. R. Astron. Soc. 425, 701–708. doi:10.1111/j.1365-2966.2012.21559.x

Longcore T., Rich C., 2004, Ecological light pollution. Frontiers in Ecology and the Environment 2, 191-198.

Navara K. J., Nelson R.J., 2007, The dark side of light at night: physiological, epidemiological, and ecological consequences. J. Pineal Res. 43, 215-224

Papoulis, A., 1981, *Systems and transforms with applications in optics*, R.E. Krieger, Florida. pp.119-128

Pun C. S. J., So C.W., 2012, Night-sky brightness monitoring in Hong Kong: A city-wide light pollution assessment. Environ Monit Assess, 184, 2537–2557. doi: 10.1007/s10661-011-2136-1

Puschnig J., Posch T, Uttenthaler S., 2014, Night sky photometry and spectroscopy performed at the Vienna University Observatory. Journal of Quantitative Spectroscopy & Radiative Transfer, 139, 64–75

Puschnig J., Schwope A., Posch T., Schwarz R., 2014b, The night sky brightness at Potsdam-Babelsberg including overcast and moonlit conditions. Journal of Quantitative Spectroscopy & RadiativeTransfer, 139, 76–81

Rabaza O., Galadí-Enríquez D., Espín Estrella A., Aznar Dols F., 2010, All-Sky brightness monitoring of light pollution with astronomical methods. Journal of Environmental Management, 91, 1278e1287

Ribas S. J., Paricio S., Canal-Domingo R., Gustems L., Calvo C. O., 2015, Monitoring Light Pollution on the Starlight Reserve of Montsec. Highlights of Spanish Astrophysics VIII, Proceedings of the XI Scientific Meeting of the Spanish Astronomical Society held on September 8-12, 2014, in Teruel, Spain. A. J. Cenarro, F. Figueras, C. Hernández-Monteagudo, J. Trujillo Bueno, and L.

Valdivielso (eds.), p. 923-928. http://adsabs.harvard.edu/abs/2015hsa8.conf..923R

Ribas S. J., Torra J., Figueras F., Paricio S., Canal-Domingo R. ,2016, How Clouds are Amplifying (or not) the Effects of ALAN. International Journal of Sustainable Lighting, 35, 32-39. doi: 10.22644/ijsl.2016.35.1.032

Ribas S. J., 2017, Caracterització de la contaminació lumínica en zones protegides i urbanes, PhD Thesis, Universitat de Barcelona, available at http://www.tdx.cat/handle/10803/396095

Rich C., Longcore T., editors, 2006, *Ecological consequences of artificial night lighting*. Washington, D.C.: Island Press.

Sánchez de Miguel A., 2016, Spatial,temporal and spectral variation of the light pollution and it sources: Methodology and results. PhD Thesis, Universidad Complutense de Madrid, http://eprints.ucm.es/31436/

Sánchez de Miguel A., Aubé M., Zamorano J., Kocifaj M., Roby J., Tapia C., 2017, Sky Quality Meter measurements in a colour changing world. Mon Not R Astron Soc, 467(3), 2966-2979. doi: /10.1093/mnras/stx145

Solano Lamphar H. A., Kocifaj M., 2016, Urban night-sky luminance due to different cloud types: A numerical experiment Lighting Res. Technol., 48, 1017–1033. doi: 10.1177/1477153515597732

Spoelstra H., Schmidt W., 2010, Lichtvervuiling boven Amsterdam (Publieksrapport)

Zamorano J., Sánchez de Miguel A., Ocaña F., Pila-Diez B., Gómez Castaño J., Pascual S., Tapia C., Gallego J., Fernandez A., Nievas M., 2016, Testing sky brightness models against radial dependency: a dense two dimensional survey around the city of Madrid, Spain. Journal of Quantitative Spectroscopy & Radiative Transfer, 181, 52–66. doi: 10.1016/j.jqsrt.2016.02.029